\documentclass[aps,twocolumn,pra,superscript,floatfix,superscriptaddress,showpacs,footinbib,reprint]{revtex4-2}
\usepackage[plainpages=false,pdfpagelabels,colorlinks=true,linkcolor=red,urlcolor=blue,
citecolor=blue,pdftitle={T-shape system v4},pdfauthor={},pdfdisplaydoctitle=true]{hyperref}
\usepackage{graphicx}
\usepackage{titlesec} 
\usepackage{float}
\usepackage{amsthm}
\usepackage{amsmath}
\usepackage{amsfonts}
\usepackage{amssymb}
\usepackage{array}
\usepackage{hhline}
\usepackage{nicefrac}
\usepackage{color}
\usepackage{subfigure}
\usepackage{dsfont}
\usepackage{txfonts}
\usepackage{wasysym}
\usepackage{multirow}
\usepackage{sidecap}
\usepackage{xcolor,cancel}
\usepackage{braket}

\begin{document}

\title{Identifying an effective model for the two-stage-Kondo regime: Numerical renormalization group results}

\author{P.~A.~Almeida}
\email{pa819824@ohio.edu}
\affiliation{Instituto de F\'isica, Universidade Federal de Uberl\^andia, 
Uberl\^andia, Minas Gerais 38400-902, Brazil}
\affiliation{Department of Physics and Astronomy and Nanoscale and Quantum Phenomena Institute,
Ohio University, Athens, Ohio 45701-2979, USA}

\author{E.~Vernek}
\affiliation{Instituto de F\'isica, Universidade Federal de Uberl\^andia, 
Uberl\^andia, Minas Gerais 38400-902, Brazil}

\author{E.~V.~Anda}
\affiliation{Departamento de F\'isica, Pontif\'icia Universidade Cat\'olica 
do Rio de Janeiro (PUC-Rio), Rio de Janeiro, Rio de Janeiro, 22453-900, Brazil}

\author{S.~E.~Ulloa}
\email{ulloa@ohio.edu}
\affiliation{Department of Physics and Astronomy and Nanoscale and Quantum Phenomena Institute,
Ohio University, Athens, Ohio 45701-2979, USA}

\author{G.~B.~Martins}
\email{gbmartins@ufu.br}
\affiliation{Instituto de F\'isica, Universidade Federal de Uberl\^andia, 
Uberl\^andia, Minas Gerais 38400-902, Brazil}

\date{\today}
\begin{abstract}
A composite impurity in a metal explores different configurations, 
where its net magnetic moment may be screened by the electrons in the host. 
An interesting example is the two-stage Kondo (TSK) system where screening sets 
in with successively smaller energy scales.  In contrast, the impurities 
may prefer a local singlet disconnected from the metal.  This competition 
is decided by fine-tuning of the couplings in the system, as has been 
studied before.  A double quantum dot T-shape geometry, where a 
`hanging’ dot is connected to current leads only via another dot, 
represents a flexible system in which these different regimes can be 
explored experimentally.  It has been difficult, however, to clearly 
differentiate the two regimes. 
Here, we provide a prescription to better identify the regime where 
the TSK occurs in such double dot geometry. The TSK regime requires a 
balance of the ratio $\nicefrac{t_{01}}{\Gamma_0}$ between 
the inter-dot coupling ($t_{01}$) and the coupling of the QD connected 
to the Fermi sea ($\Gamma_0$). Above a certain value of this ratio, the 
 system crosses over to a molecular regime, where the quantum dots form a 
 local singlet, and no Kondo screening occurs. Here, we establish that there is a region 
 in the $t_{01}\,\mbox{--}\,\Gamma_0$ parameter space where a \emph{pure} TSK 
 regime occurs, i.e., where the properties of the second Kondo stage 
 can be accurately described by a single impurity Anderson model  
 with effective/renormalized parameters. By examining the magnetic susceptibility 
 of the hanging QD, we show that a single parameter, $\Gamma_{\rm eff}$, can accurately 
 simulate this susceptibility. This \emph{effective} 
 model also provides the hanging QD spectral function with great accuracy 
 in a limited range of the $t_{01}\,\mbox{--}\,\Gamma_0$ parameter space, 
 thus defining the region where a true TSK regime occurs. We also 
 show that in this parameter range, the spin correlations between both 
 quantum dots show a universal behavior. 
 Our results may guide experimental groups to choose parameter values 
 that will place the system either in the TSK regime or in the 
 crossover to the molecular regime. 
 
\end{abstract}

\maketitle

\section{Introduction}

Point defects in semiconductors---in the form of vacancies, for example---may greatly 
affect the electrical conductivity of a host crystal~\cite{Ashcroft76}. 
Similarly, magnetic impurities 
can qualitatively alter the electrical conductivity of their 
metallic hosts at low temperatures~\cite{Dehaas1934}. The former effect, once  
understood in the 1940s~\cite{Hoddeson1992}, found its first technological applications 
in the 1950s and ushered the microelectronics revolution. The latter, 
a much more subtle effect, was first observed in the early 1930s~\cite{Dehaas1934}, 
and took three decades to be partially understood~\cite{Kondo1964} as being related 
to many-body processes. This interesting phenomenon came to be called the 
Kondo effect~\cite{Hewson1993} and it was modeled in the early 
1960s~\cite{Anderson1961} by 
the single impurity Anderson model (SIAM). It was partially solved in the mid 
1960s~\cite{Kondo1964} (through a mapping to the Kondo model~\cite{Schrieffer1966}), 
and finally solved and deeply understood at the end of the 1970s through 
renormalization group ideas~\cite{Wilson1975,Krishna-murthy1980}. 

A vacancy in a semiconductor, being a single-electron problem, does not lead to 
qualitatively different properties when a few more vacancies are added. In 
contrast, two magnetic impurities in a metallic host may interact with 
each other, even indirectly, leading to a completely different ground-state, 
depending on the ratio of the relevant parameters~\cite{Jones1988,Jones1989}. 
The phenomenology of the so-called two-impurity Anderson 
model~\cite{Sakai1990,*Sakai1992a,*Sakai1992b} lies at the heart 
of the current understanding of important classes of compounds, such as 
heavy-fermions~\cite{Stewart1984} and Kondo insulators~\cite{Tsunetsugu1997}. 

Research on single-impurity systems received a large boost at the closing 
of last century with the observation of the Kondo effect in quantum dots 
(QDs)~\cite{Goldhaber-Gordon1998} and magnetic impurities on 
metal surfaces~\cite{Kouwenhoven2001}. The ability to continuously tune 
the relevant Kondo-parameters in lithographically defined QDs opened the doors 
to detailed studies of the Kondo effect~\cite{Pustilnik2004}. Subsequently, systems 
with two QDs were built~\cite{Oosterkamp1998} and, a few years later, 
indirect coupling between two QDs was being analyzed~\cite{Craig2004,Martins2006}. 
In addition, in systems 
where there is competition between Kondo screening and other many-body interactions, 
a quantum phase transition (QPT) may occur, including the existence of non-Fermi liquid 
phases~\cite{Bulla2003,Vojta2006}. This behavior may be experimentally studied 
in multi-QD systems~\cite{Sasaki2009,Zitko2010}, in a single QD where there is orbital 
degeneracy~\cite{Zarand2006a}, in single-molecule junctions~\cite{Evers2020,Guo2021}, 
and, more recently, in coupled hybrid metal–semiconductor islands~\cite{Pouse2023}. 

\begin{widetext}

\begin{figure*}[ht]
  \centering
  \includegraphics[width=1.0\columnwidth]{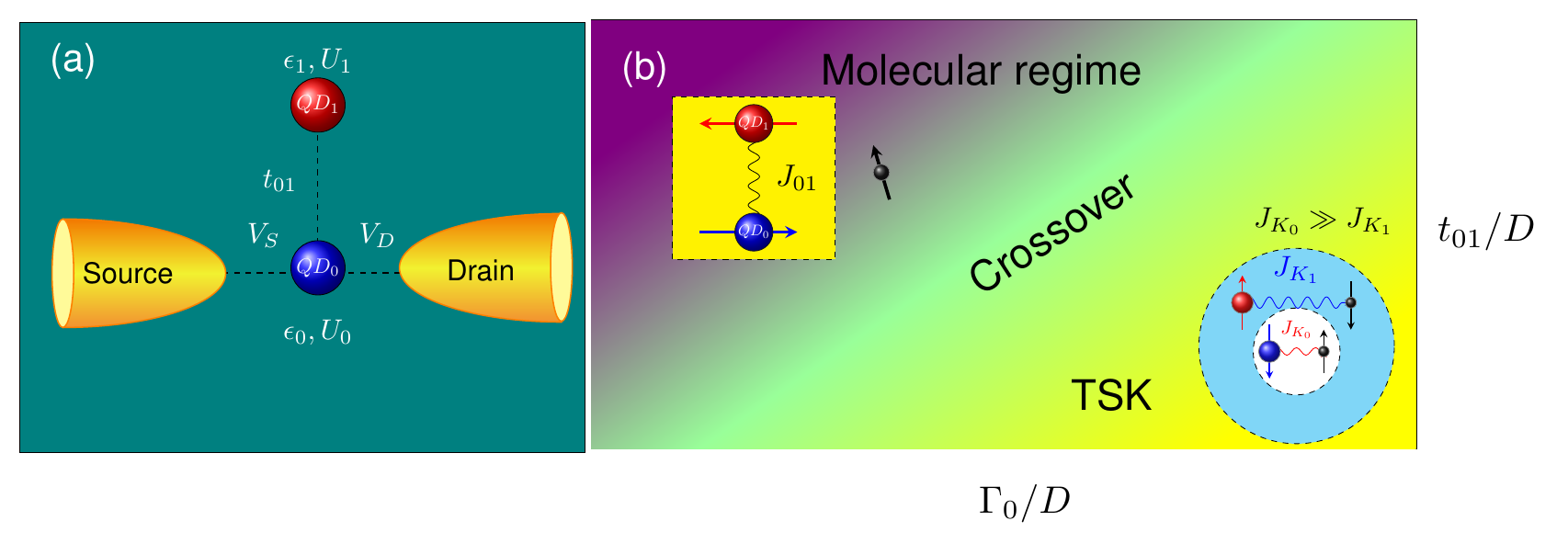}
  \caption{(a) Schematic diagram of the DQD system in T-shape geometry. Only 
        QD$_0$ is coupled to the source and drain leads through hoppings 
        $V_D$ and $V_S$. QD$_0$ and QD$_1$ are coupled by a hopping 
        parameter $t_{01}$ and their local energies are given by $\epsilon_0$ and  
        $\epsilon_1$, respectively. The on-site QD Coulomb interactions are 
        $U_0$ and $U_1$, where we take $U_0=U_1=U$, and we work in 
        the PHS point, i.e., $\epsilon_0=\epsilon_1=-U/2$. (b) A schematic 
        representation, in the $t_{01}\,\mbox{--}\,\Gamma_0$ parameter space, 
        of the two regimes present in the DQD T-shape geometry system: 
        right lower corner depicts the TSK regime, while the left 
        upper corner contains the molecular regime (see text). The crossover region 
        between these two regimes presents interesting effects 
        resulting from competing interactions between the molecular regime, 
        characterized by $J_{01} \simeq \nicefrac{4t_{01}^2}{U}$, 
        and the Kondo couplings $J_{K_1}$ and $J_{K_0}$, which control the TSK regime. 
  }
  \label{fig1}
\end{figure*}

\end{widetext}

In this work, we are interested in revisiting, using the numerical renormalization group 
(NRG) method~\cite{Wilson1975,Krishna-murthy1980,Bulla2008}, one specific double-QD (DQD) 
arrangement [see Fig.~\ref{fig1}(a)], where only one of the QDs (QD$_0$) is 
connected to the leads, while the other (QD$_1$) is side-connected to 
QD$_0$, forming a T-shape or hanging dot geometry. This system has been 
shown to exhibit the well-known two-stage Kondo (TSK) 
effect~\cite{Hofstetter2001,Vojta2002,Cornaglia2005,Zitko2006,
Zitko2007,Chung2008,Sasaki2009,Zitko2010,Tamura2010,Ferreira2011,Tanaka2012,
Baines2012,Liao2015,Crisan2015,Chen2021,Guo2021}, which may happen in a variety of systems 
where there is more than one localized magnetic moment involved, thus creating an 
interplay between local and itinerant couplings~\cite{Hofstetter2001,Vojta2002}. 
 
For the DQD system with T-shape geometry [see Fig.~\ref{fig1}(a)] in the TSK regime,  
QD$_0$ forms a Kondo state with the leads below a characteristic temperature 
$T_{K_0}$, while for $T < T_{K_1} \ll T_{K_0}$, QD$_1$ forms 
a Kondo state with the Fermi liquid resulting from the first-stage 
Kondo effect~\cite{Cornaglia2005}. Our goal is to find an effective 
\emph{single}-QD system that will 
reproduce QD$_1$'s Kondo physics (more specifically, its magnetic susceptibility 
and impurity spectral function) when QD$_1$ is in the second stage of the TSK effect. 
Based on the well-known universality of the Kondo effect, 
one expects that the Kondo physics of QD$_1$ should be 
\emph{identical} to that of an effective \emph{single} impurity Anderson 
model over a well-defined region of the $t_{01}\,\mbox{--}\,\Gamma_0$ parameter 
space with $U_{\rm eff}=U_1$, $\epsilon_{\rm eff}=\epsilon_1$, and with  
effective hybridization $\Gamma_{\rm eff}$ that is a function of $t_{01}$ 
and $\Gamma_0$. How well this effective model reproduces the Kondo properties of 
QD$_1$ can then be used to establish the range of values of $t_{01}$ 
and $\Gamma_0$ where the DQD system is indeed in a TSK regime. We will see 
that this approach is very useful in defining that range, and that, in reality, 
it is found to be quite restricted. 

As depicted in Fig.~\ref{fig1}(b), the main result obtained in this work 
is the determination of a region in the $t_{01}\,\mbox{--}\,\Gamma_0$ 
parameter space where a \emph{true} TSK regime occurs. This regime, 
located in the lower right-corner of Fig.~\ref{fig1}(b),  
lies below a critical ratio $\nicefrac{t_{01}}{\Gamma_0}$, where the 
dominance of the coupling to the leads over the interdot coupling 
allows the sequential Kondo effects 
to occur. In this regime, the dominant interactions are the Kondo interactions 
$J_{K_0}$ and $J_{K_1}$~\cite{Hewson1993}, with $J_{K_0} \gg J_{K_1}$. On the opposite 
corner, in Fig.~\ref{fig1}(b), lies the molecular regime, where the 
ratio $\nicefrac{t_{01}}{\Gamma_0}$ is such that 
$J_{01} \simeq \nicefrac{4t_{01}^2}{U}$ 
dominates, locking the QDs into a singlet, effectively disconnecting the DQD 
spins from the Fermi sea. The TSK and molecular regimes are separated by a 
crossover region, where these interactions compete.  
Finding the effective model that defines the TSK region may help design T-shape 
experimental setups to precisely tune the system into this region, and from 
there move into the crossover region and explore its properties. 

This paper is organized as follows. In Sec.~\ref{sec:Model}, we present the 
model used and the associated Hamiltonian, followed by the NRG results 
in Sec.~\ref{sec:NRG-results}, which is divided into several subsections. 
First, in subsection \ref{subsec:QD1suscep}, we define the susceptibility 
for QD$_1$, then we introduce a SIAM effective model in subsection \ref{subsec:effective}, 
which is further analyzed in subsection \ref{subsec:vGamma}. The effective 
model is then used to study the spectral function of QD$_1$ in subsection 
\ref{subsec:spec}, followed by two subsections analyzing the inter-QD spin 
correlations, subsections \ref{subsec:varyt01} and \ref{subsec:varyGamma}. 
Finally, in subsection \ref{subsec:collapse} we analyze the universality 
of the inter-QD spin correlations. Section \ref{sec:summ-con} presents a 
summary and our conclusions. 

\section{Model and Hamiltonian}\label{sec:Model}

In Fig.~\ref{fig1}(a), we show the system being analyzed in this work. 
Quantum dot QD$_0$ is connected to source and drain leads through hoppings 
$V_{S}$ and $V_{D}$, respectively, while QD$_1$ is coupled only to QD$_0$ through a hopping $t_{01}$.  
We model this system through a double-impurity Anderson model, whose  
Hamiltonian is given by $H=H_{\mathrm{DQD}}+H_{\mathrm{leads}}+H_{\mathrm{hyb}}$, 
where the first term ($H_{\mathrm{DQD}}$), describing the DQD, is  
\begin{equation}
\begin{aligned}
H_{\mathrm{DQD}}= & \sum_{i \sigma} \epsilon_i n_{i \sigma}+\sum_i U_i n_{i \uparrow} n_{i \downarrow}
	- t_{01}\sum_\sigma \left(d_{0 \sigma}^{\dagger} d_{1 \sigma}+\text { H.c. }\right),
\end{aligned}
\end{equation}
where $i=0,1$ labels the QDs, $n_{i \sigma}=d_{i \sigma}^{\dagger} d_{i \sigma}$ 
is the number operator for QD$_i$, thus $d_{i \sigma}$ annihilates an electron with spin $\sigma=\uparrow,\downarrow$ 
in QD$_i$, while $U_i$ is the Coulomb repulsion in QD$_i$ and $\epsilon_i$ is its orbital energy. Finally, 
$t_{01}$ is the amplitude for interdot hopping. The second term ($H_{\mathrm{leads}}$), describing 
the source ($r=S$) and drain ($r=D$) leads, is given by 
\begin{equation}
\begin{aligned}
H_{\mathrm{leads}}= & \sum_{\vec{k} \sigma r} \epsilon_{r \vec{k} \sigma} n_{r \vec{k} \sigma},
\end{aligned}
\end{equation}
where $n_{r \vec{k} \sigma}=c_{r \vec{k} \sigma}^{\dagger}c_{r \vec{k} \sigma}$ is the number operator 
for a state with energy $\epsilon_{r \vec{k} \sigma}$ in lead $r$. 
The third term, containing the coupling between QD$_0$ and the leads, is given by 
\begin{equation}
\begin{aligned}
H_{\mathrm{hyb}}= & \sum_{\vec{k} \sigma r}\left(V_{r \vec{k} \sigma} d_{0 \sigma}^{\dagger} c_{r \vec{k} \sigma}+\text { H.c. }\right),
\end{aligned}
\end{equation}
where $V_{r \vec{k} \sigma}$ is the (spin conserving) hopping from QD$_0$ to lead $r=S,D$. 
Taking $V_{S \vec{k} \sigma}=V_{D \vec{k} \sigma}=V_0$, QD$_0$ couples 
only to the symmetric combination of both leads, through hopping $V=\sqrt{2} V_0$; then 
the hybridization between QD$_0$ and the band is 
given by $\Gamma_0 = \pi V^2 \rho_0$, where $\rho_0$ is the lead density of states (DOS) at
the Fermi energy. For simplicity, we chose to take $V_S=V_D$, since, 
as is well known in Kondo physics, this simplifies the problem, without introducing 
any spurious effects~\cite{Hewson1993}. We also consider the metallic host as 
having a uniform (flat) DOS in the interval $-D \leq \omega \leq D$, where $D = 1$, 
half of the bandwidth, is the energy unit. In addition, for simplicity, we consider 
$U_0=U_1=U$ and $\epsilon_0=\epsilon_1=\nicefrac{-U}{2}$, 
i.e., that the system is in the particle-hole symmetric (PHS) point. 
The NRG calculations were done using the NRG Ljubljana code~\cite{zitko_rok}. 
For most of the calculations, we have used the discretization 
parameter $\Lambda=2.0$ and kept at least 5000 states at each iteration. We also employ the 
z-trick~\cite{Campo2005} (with $z=0.25$, $0.5$, $0.75$, and $1.0$, i.e., $N_z=4$) to remove 
oscillations (artifacts) in the physical quantities (for examples of its use, see Ref.~\cite{Almeida2024}). 
The thermodynamic quantities were calculated using the traditional single-shell approximation, 
while the dynamical quantities (spectral function) were calculated using the 
density matrix NRG approximation~\cite{Hofstetter2000}.

\begin{widetext}

\begin{figure*}[ht]
  \centering
  \includegraphics[width=1.0\columnwidth]{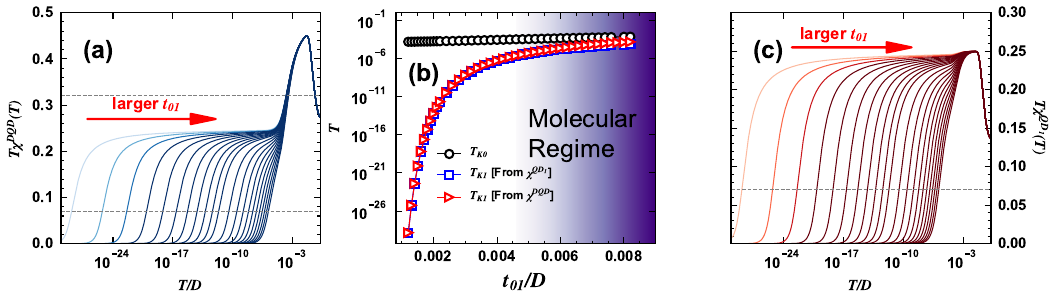}
  \caption{(a) $T~\chi^{DQD}(T)$ vs $T$ for $U=0.5$, $\Gamma_0=0.035$ 
  and $0.0012 \leq t_{01} \leq 0.0022$ in steps of $0.0001$, 
  and for $0.0024 \leq t_{01} \leq 0.0044$ in steps of $0.0002$. The red 
  arrow indicates increasing $t_{01}$ values. (b) Variation of both 
  Kondo temperatures with $t_{01}$, $T_{K_0}$ (black circles) 
  and $T_{K_1}$ (blue squares and red triangles), for the same $\Gamma_0$ as in 
  (a). As expected, $T_{K_0}$ is very weakly dependent on 
  $t_{01}$, while $T_{K_1}$ varies by $\approx 20$ orders of magnitude 
  and takes extremely low values for the smaller $t_{01}$ values. 
  The different ways of obtaining $T_{K_1}$ are discussed in the 
  text. The shading on the right side of the panel is to indicate the 
  region where the system is increasingly in a molecular regime, 
  and no longer in a TSK regime. 
  (c) $T~\chi^{QD_1}(T)$ vs $T$ for the same parameters as in panel (a). 
  The red arrow has the same meaning as in panel (a). 
  The horizontal gray dashed lines in panels (a) and (c) 
  indicate the values of the Wilson criterion parameter $\alpha_i$ 
  discussed in the text. 
  }
  \label{fig2}
\end{figure*}

\end{widetext}

\section{NRG Results}\label{sec:NRG-results}

\subsection{Obtaining $T_{K_1}$ in two different ways: defining the magnetic susceptibility for QD$_1$}\label{subsec:QD1suscep}

We use the Wilson criterion~\cite{Hewson1993} for obtaining 
$T_{K_0}$ and $T_{K_1}$, \emph{viz}., 
$T_{K_i}\,\chi^{DQD}(T_{K_i})=\alpha_i$, where 
$\alpha_0=0.25+0.07=0.32$ and $\alpha_1=0.07$. Here, $\chi^{DQD}$ is the DQD 
contribution to the total susceptibility, and it is obtained by calculating the 
total system susceptibility (including the leads) and subtracting the susceptibility 
of a reference system, i.e., the system \emph{without} the DQD (i.e., the Fermi sea 
or `bath'---see discussion 
around Eq.~(47) in Ref.~\cite{Bulla2008}). Figure \ref{fig2}(a) shows 
$T \chi^{DQD}(T)$ as a function of temperature, for $\Gamma_0=0.035$,  
in the interdot coupling range $0.0012 \leq t_{01} \leq 0.0044$. 
The two horizontal dashed lines indicate the 
$\alpha_i$ values mentioned above, used to obtain $T_{K_i}$. 
Figure \ref{fig2}(b) shows $T_{K_0}$ (black circles) and $T_{K_1}$ 
(red triangles), as a function of $t_{01}$, obtained as in 
Fig.~\ref{fig2}(a), but now for $t_{01}$ up to $0.0082$, i.e., 
very much inside the molecular regime. 
The results in panel (b) show that, as $t_{01}$ decreases, $T_{K_0}$ 
remains almost constant (as expected, since it should depend much more 
strongly on $\Gamma_0$ than on $t_{01}$), while $T_{K_1}$ decreases by 
several orders of magnitude. As it will become clearer below, for a fixed value 
of $\Gamma_0$ there is a $t_{01}$ value above which the system crosses into the 
molecular regime, and there is no more Kondo screening and 
the quantities $T_{K_0}$ and $T_{K_1}$ cease to be meaningful. The shading in 
Fig.~\ref{fig2}(b) is meant to depict the molecular regime (see Ref.~\cite{Cornaglia2005}). 

By choosing a different reference system, we may calculate the 
contribution of \emph{just} QD$_1$ to the susceptibility. Indeed, 
by subtracting, from the total system susceptibility, the 
susceptibility of the Fermi sea \emph{plus} QD$_0$ (coupled 
to the Fermi sea by $\Gamma_0$), we obtain $\chi^{QD_1}$, 
the QD$_1$ susceptibility. For the 
same parameters as in Fig.~\ref{fig2}(a), $T \chi^{QD_1}(T)$ is shown in 
Fig.~\ref{fig2}(c). Applying the Wilson criterion, with $\alpha_1=0.07$ [horizontal 
dashed line in panel (c)], to $T \chi^{QD_1}$, we obtain the blue squares 
in panel (b). The excellent agreement between the two methods [compare 
blue squares and red triangles in Fig.~\ref{fig2}(b)] confirms that 
the definition of the susceptibility of QD$_1$ is sound. 
It is the susceptibility of QD$_1$, $\chi^{QD_1}$, that will be used to 
asses the validity of a \emph{single} impurity Anderson model, 
here dubbed an \emph{effective model}, as described in the next section. 

\subsection{The effective model: fixed $\Gamma_0$ and 
varying $t_{01}$}\label{subsec:effective}

\begin{figure}
    \centering
    \includegraphics[width=1.0\columnwidth]{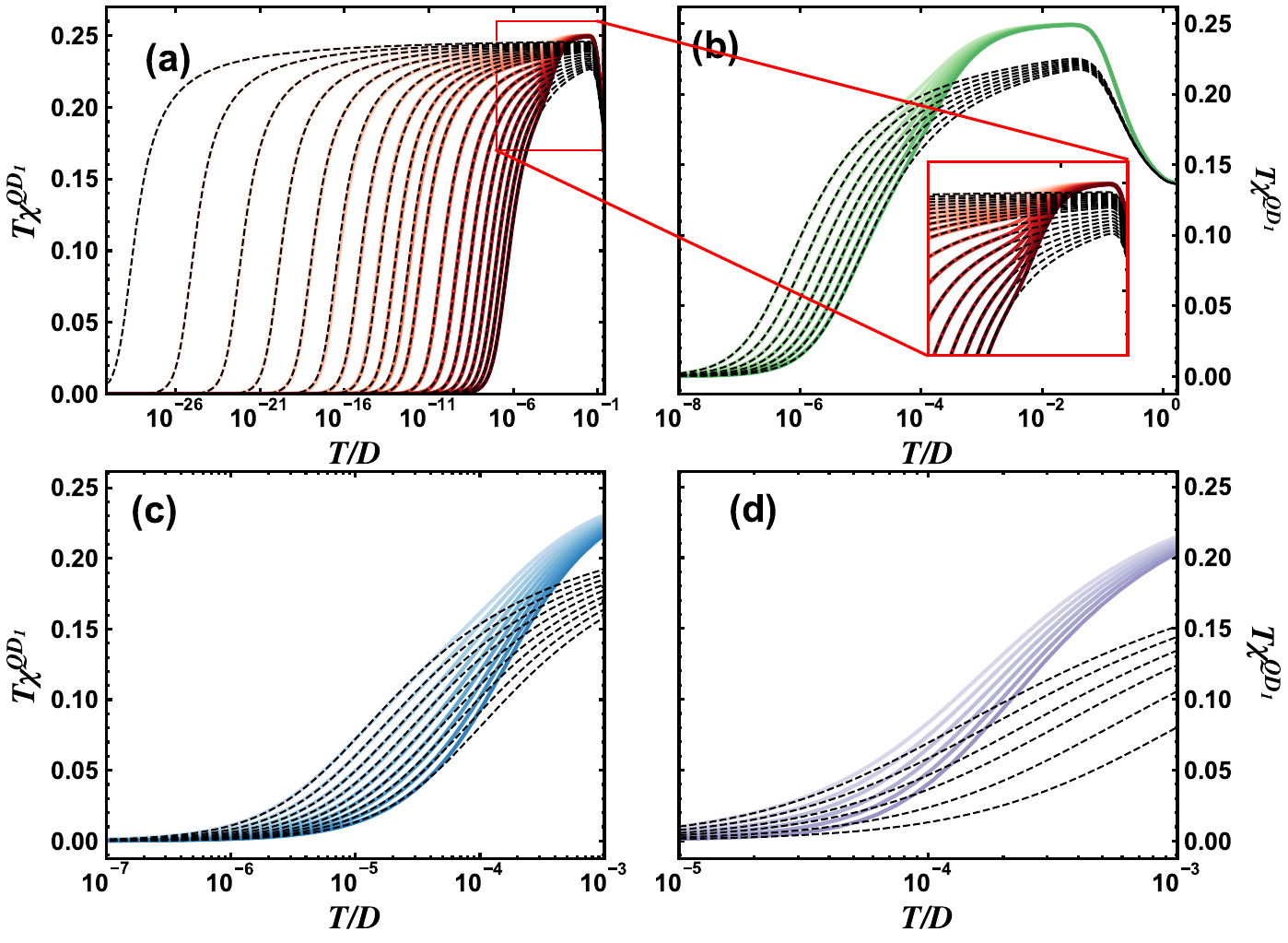}
    \caption{$T\chi^{QD_1}$ vs. $T$ results (colored solid curves) 
    compared to the results obtained through an effective SIAM (black dashed 
    curves) for different values of $t_{01}$. Parameters are 
    $U_0 = U_1= 0.5$, $\epsilon_0=\epsilon_1 = \nicefrac{-U}{2}$, and 
    $\Gamma_{0} = 0.035$. 
    (a) $0.0012 \leq t_{01} \leq 0.0022$ in steps of $0.0001$ and $0.0024 \leq t_{01} \leq 0.0038$ in 
    steps of $0.0002$; (b) $0.0040 \leq t_{01} \leq 0.0052$ in steps of $0.0002$; 
    (c) $0.0054 \leq t_{01} \leq 0.0070$ in steps of $0.0002$; 
    (d) $0.0072 \leq t_{01} \leq 0.0082$ in steps of $0.0002$. The 
    inset in panel (b) is a zoom in of the data in panel (a) of the local moment  
    fixed point temperature range. Panel (a) displays excellent agreement with 
    the effective model in that $t_{01}$ range, which dissipates for higher $t_{01}$ 
    in (b)-(d).}
    \label{fig3}
\end{figure}

Based on the universality of the Kondo effect, 
we expect that the Kondo physics of QD$_1$ should be, at least in 
some region of the $t_{01}\,\mbox{--}\,\Gamma_0$  parameter space, 
\emph{identical} to that of an effective SIAM with $U_{\rm eff}=U_1$, 
$\epsilon_{\rm eff}=\epsilon_1=\nicefrac{-U}{2}$, and with an 
effective hybridization $\Gamma_{\rm eff}$ that should depend on $t_{01}$ 
and $\Gamma_0$. Indeed, Fig.~\ref{fig3}(a) shows the 
fitting (black dashed curves) of $T\chi^{QD_1}$ 
(solid color curves) for the interval $0.0012 \leq t_{01} 
\leq 0.0038$. By appropriately adjusting the 
value of $\Gamma_{\rm eff}$ (which functions as a fitting parameter), for each 
different value of $t_{01}$ (for $\Gamma_0=0.035)$, we are able 
to accurately reproduce the $T\chi^{QD_1}$ results (color curves) 
in Fig.~\ref{fig3}(a) over a range of $t_{01}$ values. This is shown by 
the dashed black curves in 
Fig.~\ref{fig3}(a), which faithfully reproduce the $T\chi^{QD_1}$ up 
to temperatures several orders of magnitude higher than $T_{K_1}$~\cite{ba}. 
As $t_{01}$ increases [panels (b) to (d)], the fitting is accurate only 
up to lower and lower temperatures, 
since the system is crossing over to the molecular regime, 
where the QDs form a \emph{local} singlet among themselves, 
and not even the first Kondo stage occurs. It is interesting to 
contrast what happens in panel (a) (TSK regime), around the 
local moment (LM) fixed point, for QD$_1$ and the 
effective model [see the inset 
in panel (b)]: since QD$_1$ is not connected directly to the Fermi 
sea, it undergoes very small charge fluctuations, thus its 
susceptibility is very close to $1/4$ for all $t_{01}$ values 
(colored solid curves), while it decreases considerably as $\Gamma_{\rm eff}$ 
increases for the effective SIAM (black dashed curves). 

\begin{figure}
    \centering
    \includegraphics[width=1.0\columnwidth]{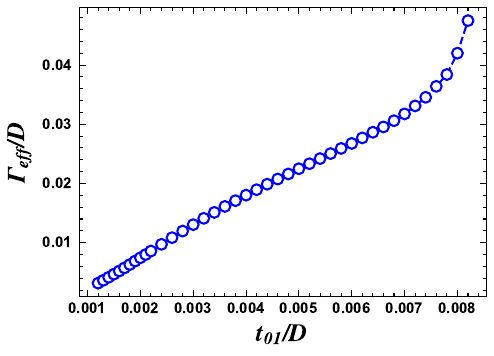}
    \caption{$\Gamma_{\rm eff}$ vs $t_{01}$ results (blue circles), 
    as obtained through the fittings done in Fig.~\ref{fig3}. 
    Notice the initial near linear dependence of $\Gamma_{\rm eff}$ 
    on $t_{01}$.}
    \label{fig4}
\end{figure}

Figure \ref{fig4} presents the $\Gamma_{\rm eff}$ values as a function of 
$t_{01}$. It is interesting to see that the region for smaller values 
of $t_{01}$ where one expects the TSK to occur~\cite{Cornaglia2005}, presents 
an almost linear dependence of $\Gamma_{\rm eff}$ with $t_{01}$. 

\subsection{QD$_1$ susceptibility for fixed $t_{01}$ and 
varying $\Gamma_0$}\label{subsec:vGamma}

\begin{figure}
    \centering
    \includegraphics[width=1.0\columnwidth]{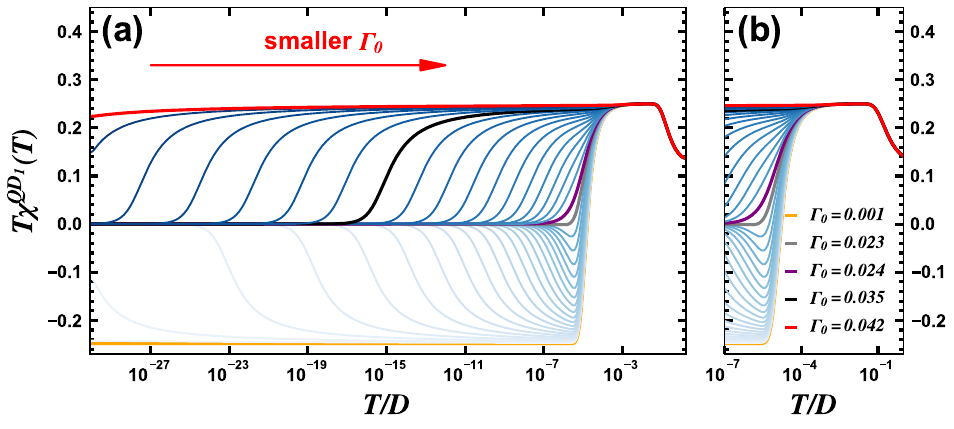}
    \caption{(a) $T\chi^{QD_1}$ vs $T$, for 
    $0.002 \leq \Gamma_0 \leq 0.042$. The system is in the PSH point, $U= 0.5$, 
    and $t_{01} = 0.0018$. The red arrow indicates the direction of decreasing $\Gamma_0$. 
    (b) Zoom in, around the LM temperature interval, of the data in panel (a). 
    The legend applies to both panels. 
    }
    \label{fig5}
\end{figure}

\begin{figure}
    \centering
    \includegraphics[width=1.0\columnwidth]{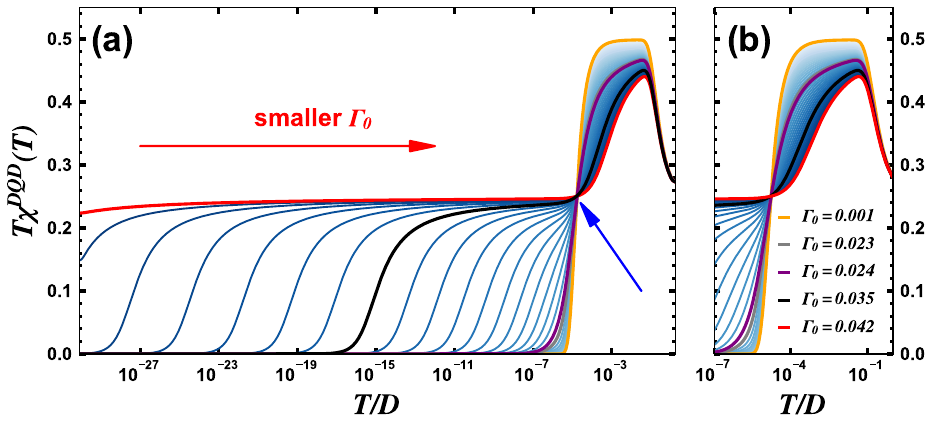}
    \caption{(a) $T\,\chi^{DQD}$ vs $T$ for several $\Gamma_0$ values. 
    Same parameters and same colors as in Fig.~\ref{fig5}. The red arrow points 
    in the direction of decreasing $\Gamma_0$ curves. The 
    blue arrow points to the peculiar crossing of all curves 
    in a single point. (b) Zoom in, around the LM temperature interval, 
    of the data in panel (a). The legend applies to both panels.
    }
    \label{fig6}
\end{figure}

We now analyze $\chi^{QD_1}$ for varying coupling to the leads $\Gamma_0$ and a fixed 
$t_{01}$ value. To help interpret these results, one should note 
that $T\chi$ (if we take a unit value for the giromagnetic factor 
$g$, Bohr magneton $\mu_B$, and Boltzmann constant $k_B$) 
measures $S_z^2$, i.e., the square of the $z$-component of the magnetic 
impurity spin. Thus, $\chi^{DQD}$ measures $S_{DQD,z}^2$ and 
$\chi^{QD_1}$ measures $S_{1,z}^2$, where 
$\vec{S}_{DQD}=\vec{S}_0 + \vec{S}_1$. 
The left panel in Fig.~\ref{fig5} shows $T\chi^{QD_1}$ as a 
function of temperature, in the PHS point, for 
$0.002 \leq \Gamma_0 \leq 0.042$, $t_{01}=0.0018$, and $U=0.5$. 
The red curve is for $\Gamma_0=0.042$ (largest $\Gamma_0$ and 
lowest $T_{K_1}$), where $T_{K_1}$ increases as $\Gamma_0$ decreases 
(the red arrow points in the direction of decreasing $\Gamma_0$ curves). 
The black curve is for $\Gamma_0=0.035$, 
resulting in $T_{K_1}=1.05 \times 10^{-15}$. Figure \ref{fig5}(b) shows a zoom 
of the LM fixed point temperature region. It is interesting to 
note that, for temperatures just below the LM temperature region 
(marked by the $S_{1,z}^2=\nicefrac{1}{4}$ plateau), the QD$_1$ susceptibility 
for $\Gamma_0 = 0.023$  (gray curve) becomes slightly negative, 
before vanishing. For progressively smaller $\Gamma_0$ values, the 
susceptibility will dip into more and more negative values, before 
going through an upturn and then vanishing as $T \rightarrow 0$. For even smaller 
$\Gamma_0$ values, the dip approaches $\nicefrac{-1}{4}$, forming a plateau at 
this value, eventually upturning and vanishing. 
This behavior reflects the procedure used to obtain $\chi^{QD_1}$ as follows. 
For $0.001 \leq \Gamma_0 \leq 0.023$ (the smaller $\Gamma_0$ values) and 
$t_{01}=0.0018$, the system is in the molecular regime (as will be shown in 
Fig.~\ref{fig11}). There, for temperatures just below the LM fixed point, 
the DQD system has $S_{DQD,z}^2 \approx 0$, because of the inter-QD singlet formation 
(see Fig.~\ref{fig6}). However, since we subtract the susceptibility of QD$_0$ 
(our reference in the $\chi^{QD_1}$ calculation), which, for this temperature and 
$\Gamma_0$ interval, is of the order of $S_{0,z}^2 \lesssim \nicefrac{1}{4}$, 
the result for $T \chi^{QD_1}$ must become negative and then 
(as $\Gamma_0$ decreases) form a plateau at $\approx \nicefrac{-1}{4}$. 
For temperatures below $T_{K_0}$, when $S_{0,z}^2$ (the reference) 
vanishes (due to Kondo), then $S_{1,z}^2-S_{0,z}^2$ also vanishes 
(since both are $\approx 0$). Thus, this calculation of the susceptibility 
for QD$_1$ for the smaller $\Gamma_0$ values is revealing. It 
is able to clearly spot the molecular regime (the onset of negative 
values for $T\,\chi^{QD_1}$), an effect that is not as apparent when we 
analyze $T\,\chi^{DQD}$ (Fig.~\ref{fig6}). Note that the results in 
Fig.~\ref{fig5} indicate that, for $\Gamma_0 \approx 0.023$ 
(and $t_{01}=0.0018$) the system is in the final stage of the crossover 
into the molecular regime. 

Figure \ref{fig6} presents results for $T\,\chi^{DQD}$, as 
a function of temperature, for the same parameters as 
in Fig~\ref{fig5}. It is interesting to note that 
there is one specific temperature at which the susceptibility 
is the same ($T\chi^{DQD} \approx \nicefrac{1}{4}$) for 
all $\Gamma_0$, indicated by a 
blue arrow. This energy scale is of the same order of magnitude as 
that signaling the crossover into the molecular regime, $\approx 10^{-5}$. 
We redid the calculations in Fig.~\ref{fig6} for different $t_{01}$ 
values, and obtained that the crossing still occurs at 
$T\chi^{DQD} \approx \nicefrac{1}{4}$, albeit at slightly different 
temperatures, while the crossing has no dependence on $\epsilon_d$.

\subsection{QD$_1$ spectral function and the effective SIAM}\label{subsec:spec}

\begin{figure}
    \centering
    \includegraphics[width=1.0\columnwidth]{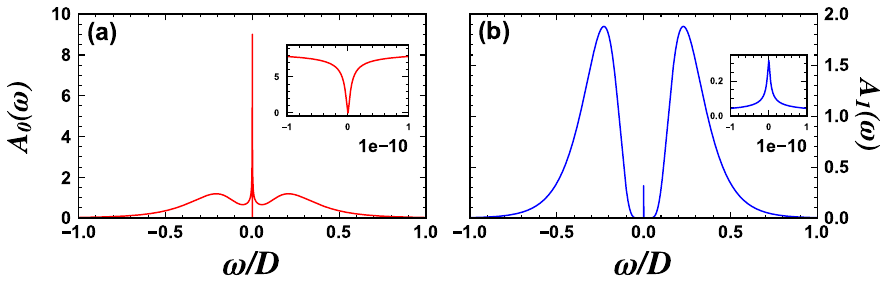}
    \caption{(a) QD$_0$ spectral function $A_0(\omega)$ for 
    $\Gamma_0=0.035$, $U=0.5$, and $t_{01}=0.0022$, in the PHS 
    point. The inset shows the antiresonance `inside' the Kondo 
    peak. (b) QD$_1$ spectral function $A_1(\omega)$ for the 
    same parameters as in (a). The inset shows a zoom of the 
    very small Kondo peak in QD$_1$. The spectral functions in 
    both panels integrate to unity. 
    } 
    \label{fig7}
\end{figure}

As shown in the previous two subsections, the effective model 
provides an excellent description of the magnetic susceptibility of 
QD$_1$, yields a good understanding of the physics around 
the LM fixed point [see inset in Fig.~\ref{fig3}(b)], and 
affords a quantitative way of determining when the system enters 
the molecular regime. Then, one may ask whether the effective model 
can describe the dynamic properties of QD$_1$. Figure \ref{fig7} shows 
the spectral functions $A_0(\omega)$ and $A_1(\omega)$, for QD$_0$ 
and QD$_1$, in panels (a) and (b), respectively. The spectral 
functions were calculated using the density matrix NRG 
approximation~\cite{Hofstetter2000}. It is well known~\cite{Cornaglia2005} that 
an antiresonance appears at the Fermi energy in $A_0(\omega)$, with a 
characteristic width $T_{K_1}$. This has been well studied in the literature and it is 
shown in the inset to Fig.~\ref{fig7}(a). However, $A_1(\omega)$ has received 
much less attention~\cite{Liao2015}. Its peculiar shape [blue curve 
in Fig.~\ref{fig7}(b)], with a Kondo 
peak much shorter than the Coulomb blockade peaks, has its origin 
in the fact that the Friedel Sum Rule for the DQD system has a much 
different expression from the well-known one for the 
SIAM~\cite{Hewson1993}, as illustrated in the case of a two-level QD 
system~\cite{Logan2009}, for example. Because of that, it is clear 
that $A_{\rm eff}(\omega)$ will not simulate $A_1(\omega)$ for an 
$\omega$-range that includes the Coulomb blockade peaks. 

\begin{figure}
    \centering
    \includegraphics[width=1.0\columnwidth]{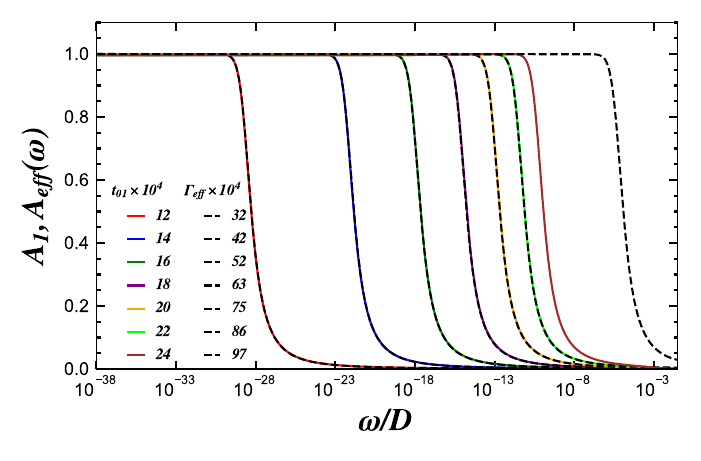}
    \caption{Comparison between the spectral functions 
    $A_1(\omega)$ (colored solid curves) and $A_{\rm eff}(\omega)$ (dashed 
    black curves) in the range $10^{-38} \leq \nicefrac{\omega}{D} \lesssim 10^{-2}$ 
    for several values of $t_{01}$ (as indicated in the legend). 
    Parameters used are $0.0012 \leq t_{01} \leq 0.0024$, $U = 0.5$, 
    $\Gamma_{0} = 0.035$, and the system is in the PHS point. The corresponding 
    values of $\Gamma_{eff}$ are indicated. Note that for $t_{01}=0.0024$ there 
    is no agreement between $A_1(\omega)$ (solid curve) and 
    $A_{\rm eff}(\omega)$ (dashed curve).}
    \label{fig8}
\end{figure}

Thus, as shown in Fig.~\ref{fig8}, we limit the comparison 
of the spectral functions to an $\omega$-range around the 
Fermi energy that excludes the Hubbard peak. We normalize to $1$ 
the height of both Kondo peaks, the one obtained from 
the effective model (black dashed curves)~\cite{normalization}
and the one from the DQD (colored solid curves), 
to facilitate comparison. It is important to stress that no 
further adjustments are needed to the effective model. In other 
words, the values of $\Gamma_{\rm eff}$ used to obtain $A_{\rm eff}(\omega)$ 
are the ones obtained from the susceptibility fittings in Fig.~\ref{fig3}, and 
shown explicitly in Fig.~\ref{fig4}~\cite{frota}. The spectral function comparisons are 
shown in Fig.~\ref{fig8}, for $\Gamma_0=0.035$ and $t_{01}$ in the interval 
$0.0012 \leq t_{01} \leq 0.0024$, as indicated in the legend. 
The agreement between $A_{1}(\omega)$ (colored solid curves) and  
$A_{\rm eff}(\omega)$ (black dashed curves), up to $t_{01}=0.0022$ 
is remarkable. For $t_{01} \geq 0.0024$, the Kondo peaks for QD$_1$ and 
the effective model deviate from each other (as shown for $t_{01}=0.0024$). 
These results suggest a different criterion to determine the range of 
$t_{01}$ values where the second Kondo effect can be truly 
simulated by the SIAM. The spectral function comparisons 
indicate that the upper limit for $t_{01}$ is considerably smaller 
than what transpires from the $T\,\chi^{QD_1}$ fittings in 
Fig.~\ref{fig3}(a), and what has been often determined in 
the literature as the TSK-regime range~\cite{Cornaglia2005}.

In the next subsections, we will study some aspects of the inter-QD spin 
correlation $\langle \vec{S}_0 \cdot \vec{S}_1\rangle$ and show 
that its variation with the system parameters ($\Gamma_0$ and $t_{01}$) may 
be used to further characterize the TSK regime. 

\subsection{Inter-QD spin correlations: Dependence on $t_{01}$}\label{subsec:varyt01}

\begin{figure}
    \centering
    \includegraphics[width=1.0\columnwidth]{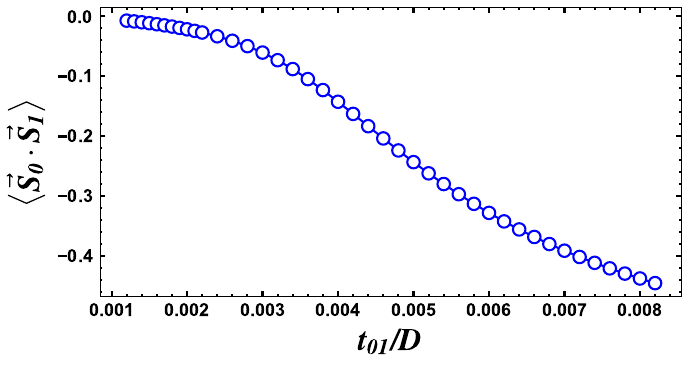}
    \caption{Ground state ($T \simeq 10^{-27}$) interdot spin correlations. 
    $\langle \vec{S}_0 \cdot \vec{S}_1\rangle$ vs $t_{01}$ for $U = 0.5$, 
    $\Gamma_{0} =  0.035$, with the system in the PHS point.
}
    \label{fig9}
\end{figure}

Figure \ref{fig9} presents $\langle \vec{S}_0 \cdot \vec{S}_1\rangle$ 
for $0.0012 \leq t_{01} \leq 0.0082$, $\Gamma_0=0.035$, $U=0.5$, 
$T \simeq 10^{-27}$ (ground state), for the system in the PHS point. Three 
different regions may be discerned. For very small $t_{01}$ values, when the system 
is in the TSK regime and thus each QD is `locked' into its own Kondo 
state, there is very little correlation between the QDs. On the other 
hand, for the largest $t_{01}$ values (molecular regime), the first 
Kondo stage does not occur (thus, neither the second), since the two 
QDs are strongly antiferromagneticaly linked by its correlation, which 
tends to the full singlet value of \nicefrac{-3}{4}. Then, in an 
intermediate region of $t_{01}$ values, 
there is a crossover between the TSK and the molecular regimes. 

\begin{figure}
    \centering
    \includegraphics[width=1.0\columnwidth]{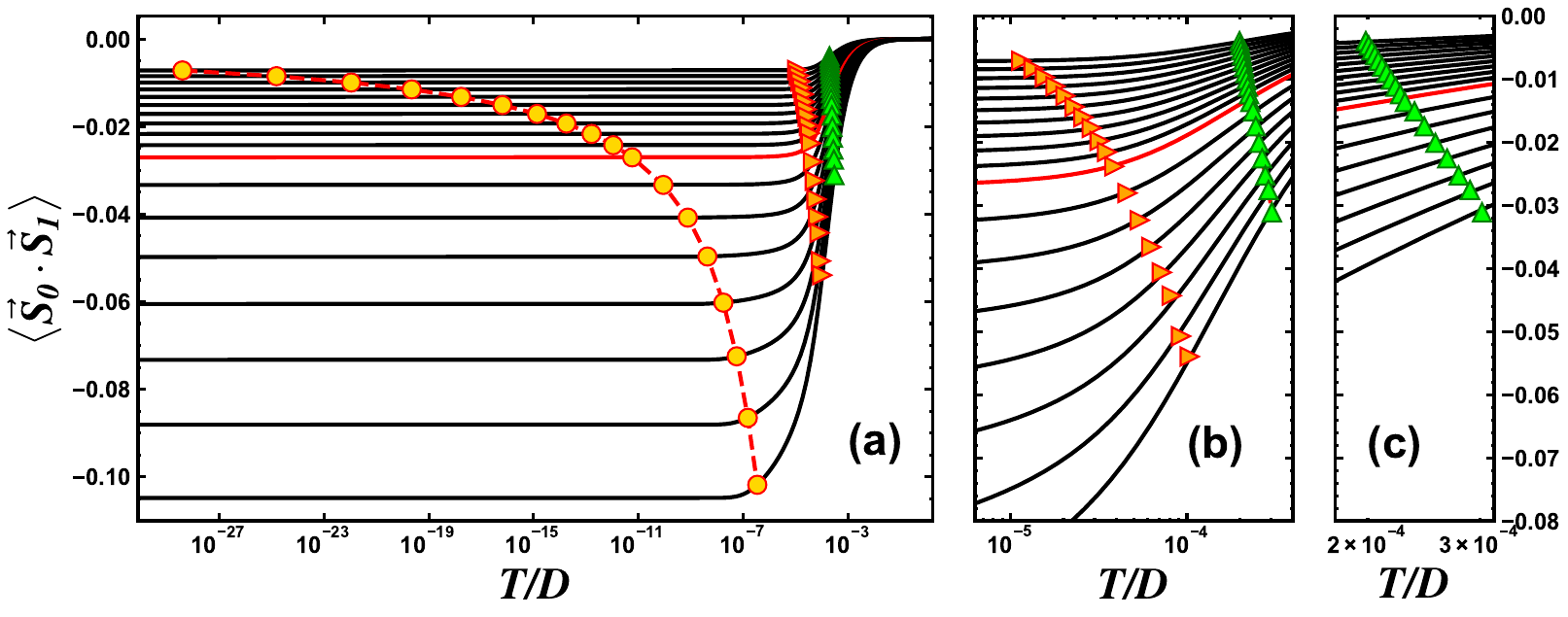}
    \caption{(a) $\langle \vec{S}_0 \cdot \vec{S}_1 \rangle$
  vs $T$ for $0.0012 \leq t_{01} \leq 0.0022$, 
  in steps of $0.0001$, and for $0.0022 \leq t_{01} \leq 0.0036$, in 
  steps of $0.0002$. Parameters used are $U = 0.5$, 
  $\Gamma_{0} =  0.035$, and system in the PHS point. 
  The red curve is for $t_{01}=0.0022$. 
  The yellow circles indicate $T_{K_1}$, the orange triangles 
  $J_{01}=\nicefrac{4t_{01}^2}{U}$, and the green triangles $T_{K_0}$. 
  (b) and (c) are progressive zoom ins on the data in panel (a) 
  at higher temperatures.
  }
    \label{fig10}
\end{figure}

It is interesting to study $\langle \vec{S}_0 \cdot \vec{S}_1\rangle$ as 
a function of temperature in the interval $0.0012 \leq t_{01} 
\leq 0.0036$, for $\Gamma_0=0.035$, $U=0.5$, in the PHS point. 
This is shown in Fig.~\ref{fig10}(a), also 
with values of $T_{K_1}$ (yellow circles), $T_{K_0}$ (green up-triangles), 
and $J_{01}=4t_{01}^2/U$ (orange right-triangles, representing the effective 
antiferromagnetic (AF) coupling between the two QDs). The symbols 
for each of these three energy scales, which depend on $t_{01}$, 
are placed on top of the corresponding $\langle \vec{S}_0 \cdot 
\vec{S}_1\rangle$ vs $T$ curve. The red curve indicates $t_{01}=0.0022$, 
which marks the end of the TSK regime, according to the spectral 
function analysis of the effective model. At high temperatures, 
the correlations vanish for all $t_{01}$ values, as expected. 
As the temperature decreases, AF correlations between the 
QDs start to develop, and they are enhanced (becoming more 
negative) even for temperatures below $T_{K_0}$ [marked by the 
green up-triangles, see zoom ins in panels (b) and (c)], 
indicating that QD$_0$ develops singlet correlations 
simultaneously with the conduction electrons (forming the Kondo 
singlet) and with QD$_1$, mainly when in the crossover regime 
($t_{01} > 0.0022$). Indeed, inside the crossover regime, 
$0.0022 < t_{01} \leq 0.0036$, the singlet correlations 
between the two QDs are strongly enhanced, for temperatures below 
$T_{K_0}$, finally settling into a plateau. This can 
be seen as a precursor to the full molecular regime. 

Now, let us analyze the onset of the second Kondo stage for the larger values 
of $t_{01}$. It is clear that down to $t_{01} \approx 0.003$ [fourth 
curve from bottom to top in panel (a)], $T_{K_1}$ marks the temperature 
where the plateau in $\langle \vec{S}_0 \cdot \vec{S}_1\rangle$ starts. 
In contrast, for the smallest values of $t_{01}$, as shown in 
panel (b), it is the energy scale $J_{01}$ (orange right-triangles) 
that marks the onset of the $\langle \vec{S}_0 \cdot \vec{S}_1\rangle$ 
plateau. For intermediate values of $t_{01}$, the plateau onset occurs 
for an energy scale that is intermediate between $J_{01}$ and $T_{K_1}$. Note that 
this analysis, for $t_{01} > 0.0036$ (for $\Gamma_0=0.035$), 
becomes questionable, since the very use of the $T_{K_0}$ and 
$T_{K_1}$ energy scales loses its meaning. 

One interesting aspect of the results in Fig.~\ref{fig10}(a) is 
that for temperatures below $T_{K_1}$ (yellow circles), there 
is virtually no change in $\langle \vec{S}_0 
\cdot \vec{S}_1 \rangle$ as the QD$_1$ spin is screened by 
the conduction electrons via the formation of the second Kondo state. 
This is especially true inside the TSK regime, where the plateau 
starts at temperatures well above $T_{K_1}$. 
In other words, the formation of a correlated state between QD$_1$ 
and the conduction electrons (the second Kondo state) does not 
alter the spin correlation of QD$_1$ with QD$_0$. 

\subsection{Inter-QD spin correlations: Dependence on $\Gamma_0$}
\label{subsec:varyGamma}

\begin{figure}
    \centering
    \includegraphics[width=1.0\columnwidth]{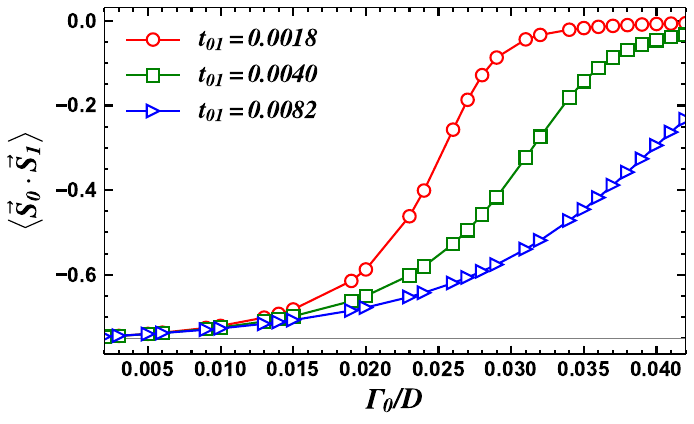}
    \caption{Ground state $\langle\vec{S}_0 \cdot \vec{S}_1\rangle$ vs. $\Gamma_0$ 
    for $t_{01}=0.0018$ (red circles), $t_{01}=0.0040$ (green squares), 
    and $t_{01}=0.0082$ (blue triangles) for $0.002 \leq \Gamma_0 \leq 0.042$, 
    $U = 0.5$, and system in the PHS point. The horizontal light-gray line 
    indicates the full-singlet value $\langle\vec{S}_0 \cdot \vec{S}_1\rangle = -3/4$. 
    }
    \label{fig11}
\end{figure}

Figure \ref{fig11} shows how $\langle \vec{S}_0 \cdot \vec{S}_1 \rangle$ 
varies with $\Gamma_0$ for three different $t_{01}$ values, \emph{viz}., 
$0.0018$, $0.0040$, and $0.0082$, for $T \simeq 10^{-27}$ (ground state). 
For very small values of $\Gamma_0$ (up to $\approx 0.015$) the QDs 
form a strong singlet ($\langle \vec{S}_0 \cdot \vec{S}_1 \rangle 
\approx -0.75$, indicated by the solid gray horizontal line), 
independent of the value of $t_{01}$. Thus, 
all three $t_{01}$ values place the system in the molecular regime 
for such small $\Gamma_0$ values. Larger $\Gamma_0$ values 
($\Gamma_0 \geq 0.025$), show a clear 
separation between the three curves: for the lowest $t_{01}=0.0018$ 
(red circles), $\langle\vec{S}_0 \cdot \vec{S}_1\rangle$ rapidly 
vanishes with increasing $\Gamma_0$, while for $t_{01}=0.0040$ 
(green squares), $\langle\vec{S}_0 \cdot \vec{S}_1\rangle$ is still strongly 
AF, but approaches zero as $\Gamma_0$ increases, as it will eventually 
crossover into the TSK regime. The $\langle\vec{S}_0 \cdot \vec{S}_1\rangle$ 
results for $t_{01}=0.0082$ (blue triangles), however, display sizeable AF 
values for nearly the whole $\Gamma_0$ interval in Fig.~\ref{fig11}, 
have a markedly different dependence on $\Gamma_0$, and become negligible 
only for $\Gamma_0$ values large enough to place QD$_0$ close to (or into) 
an intermediate valence regime~\cite{Hewson1993}. 

Coming back to the $t_{01}=0.0018$ results (red circles), the 
spectral function in Fig~\ref{fig8}(d) shows that for 
$\Gamma_0=0.035$ the system is well into the TSK regime, since the 
QD$_1$ dynamical properties agree with the SIAM effective model. 
Obviously, larger values of $\Gamma_0$ will maintain the system in 
the TSK regime. The functional form of the red circles curve in Fig.~\ref{fig11} 
seems related to that of the other two curves, which motivates one to explore 
re-scaling properties. This probing for universal behavior is carried out 
in the next subsection. 

\subsection{Collapse of the inter-QD spin correlations}\label{subsec:collapse}

\begin{widetext}

\begin{figure*}[ht]
    \centering
    \includegraphics[width=1.0\columnwidth]{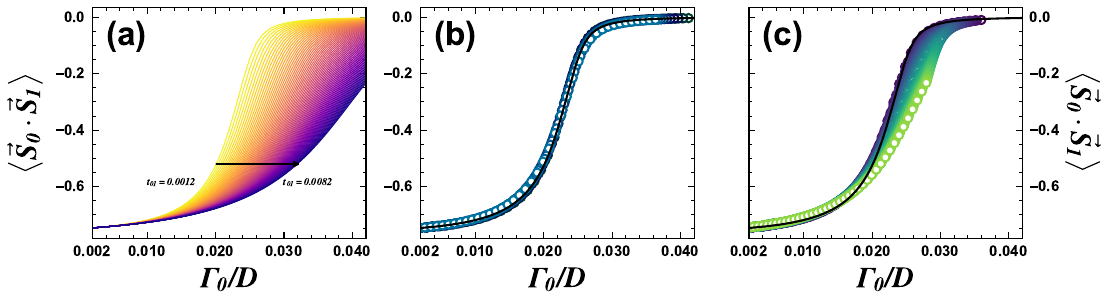}
    \caption{(a) Ground state $\langle\vec{S}_0 \cdot \vec{S}_1\rangle$ vs $\Gamma_0$ for 
    $0.0012 \leq t_{01} \leq 0.0082$ (from left to right, see horizontal arrow). 
    (b) Re-scaled $0.0012 \leq t_{01} \leq 0.0023$ results, showing their  
    collapse onto the $t_{01}=0.0012$ curve (black solid line). 
    (c) $\langle\vec{S}_0 \cdot \vec{S}_1\rangle$ results for the interval 
    $0.0024 \leq t_{01} \leq 0.0082$, showing that these results do not share 
    universality with the lower $t_{01}$ results in panel (b).
    }
    \label{fig12}
\end{figure*}

\end{widetext}

Figure \ref{fig12}(a) shows the $\langle\vec{S}_0 \cdot \vec{S}_1\rangle$ 
vs $\Gamma_0$ curves in the interval $0.0012 \leq t_{01} \leq 0.0082$. 
In accordance to the discussion above, we use the lowest $t_{01}=0.0012$ 
curve in Fig.~\ref{fig12}(a) as the `target' 
curve, and try to collapse higher $t_{01}$-value curves onto it. 
%In Fig.~\ref{fig12}(b), we show an interpolation (black solid curve) of 
%the $t_{01}=0.0012$ results (blue circles). 
We apply the following procedure. 
First, we choose a `reference' value, denoted $\Gamma_0^{\rm ref}$, 
inside the TSK region of the target curve. 
Next, we find what is the $\Gamma_0$ value (denoted $\Gamma_0^{0.0012}$) 
for which $\langle\vec{S}_0 \cdot \vec{S}_1\rangle(t_{01}=0.0012,\Gamma_0^{0.0012})=
\langle\vec{S}_0 \cdot \vec{S}_1\rangle(t_{01},\Gamma_0^{\rm ref})$, providing 
the re-scaling factor by which we divide the $\Gamma_0$ axis for the 
$t_{01}$ curve in question, $\xi = \Gamma_0^{\rm ref}/\Gamma_0^{0.0012}$. 
Panel (b) shows the re-scaled results in the interval $0.0013 \leq t_{01} 
\leq 0.0023$, using $\Gamma_0^{\rm ref}=0.027$, where all curves collapse 
onto the curve for $t_{01}=0.0012$ (black solid curve). 
Panel (c) shows that applying the same re-scaling procedure for $t_{01}>0.0023$ 
curves does not produce as good a collapse onto the target curve. 

It should be noted that the $\langle\vec{S}_0 \cdot \vec{S}_1\rangle$ 
curves for larger values of $t_{01}$ [see panel (c)] may re-scale for  
larger values of $\Gamma_0^{\rm ref}$, at least until a certain value of $t_{01}$. 
This reflects the fact that it is the ratio $\nicefrac{t_{01}}{\Gamma_0}$ 
that determines if the system is in the TSK regime or not (as long as $\Gamma_0$ 
does not result in a mixed valence regime). 

\begin{figure}
    \vspace{0.3cm}
    \centering
    \includegraphics[width=1.0\columnwidth]{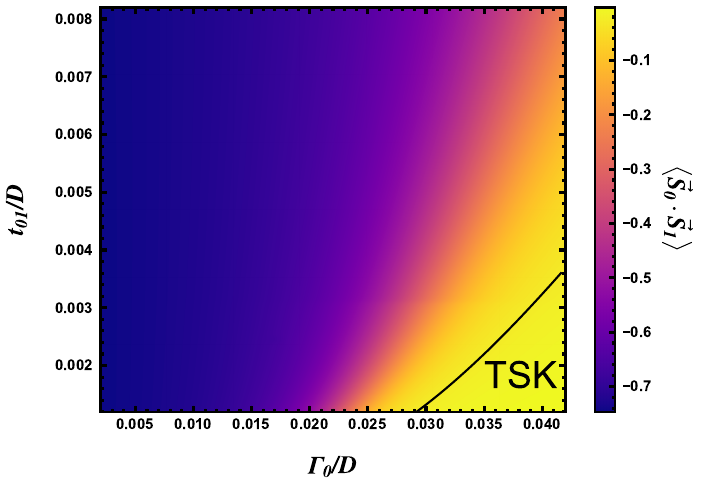}
    \caption{$\langle\vec{S}_0 \cdot \vec{S}_1\rangle$ color map in the 
    $t_{01}\,\mbox{--}\,\Gamma_0$ parameter space. The solid line is  
    defined as $\langle\vec{S}_0 \cdot \vec{S}_1\rangle(t_{01},\Gamma_0) = 
    -0.027$ and delimits the location of the TSK regime, as defined 
    in this work. 
    }
    \label{fig13}
\end{figure}

Figure \ref{fig13} shows $\langle\vec{S}_0 \cdot \vec{S}_1\rangle$ as a color 
map in the $t_{01}\,\mbox{--}\,\Gamma_0$ parameter space. We can use it 
to delineate a region where the TSK regime 
resides. For that, we use the spectral function results 
obtained previously, where we found the maximum $t_{01}=0.0022$ 
(for $\Gamma_0=0.035$) for which the QD$_1$ Kondo peak could be accurately 
simulated through the effective model. For these values we 
obtain $\langle\vec{S}_0 \cdot \vec{S}_1\rangle(0.0022,0.035)=-0.027$. 
Taking that as the lower limit for $\langle\vec{S}_0 \cdot \vec{S}_1\rangle$, that 
defines the border of the TSK regime, we obtain the slanted black line 
close to the lower right corner in Fig.~\ref{fig13}. In other words, this 
is the line where $\langle\vec{S}_0 \cdot \vec{S}_1\rangle(t_{01},\Gamma_0)=-0.027$ 
in the $t_{01}\,\mbox{--}\,\Gamma_0$ plane. The yellow region, 
for which $\langle\vec{S}_0 \cdot \vec{S}_1\rangle(t_{01},\Gamma_0) > -0.027$, 
defines the TSK regime. 

\section{Summary and Conclusions}\label{sec:summ-con}

In summary, we have analyzed a T-shape geometry for a DQD system, using NRG. 
Our simulation through a two-impurity Anderson model is focused on determining 
the $t_{01}\,\mbox{--}\,\Gamma_0$ parameter region in which the second 
Kondo stage can be well defined. We were able to show, using an effective 
SIAM, that the TSK regime occupies a rather restricted 
sector of parameter space (see Fig.~\ref{fig13}). The process of obtaining 
this result required the calculation of the QD$_1$ susceptibility, its 
comparison with the susceptibility of an effective SIAM, as well as the 
comparison of the dynamic properties of QD$_1$. We found that in the 
region where the ``better-defined'' TSK occurs (for fixed 
$\Gamma_0$ and small values of $t_{01}$), $\Gamma_{\rm eff}$ depends 
linearly on $t_{01}$ (see Fig.~\ref{fig4}). 

To supplement the information and perspective that $\chi^{QD_1}$ and $A_1(\omega)$ 
can provide, we also carried out a thorough analysis of the inter-QD spin correlations 
$\langle\vec{S}_0 \cdot \vec{S}_1\rangle$. This allowed us 
to uncover interesting features of the DQD system, \emph{viz}., the TSK 
manifests itself through the appearance of a plateau in the interdot spin correlation 
for $T < J_{01}$ (see Fig.~\ref{fig10}) and through 
the universality of the $\langle\vec{S}_0 \cdot \vec{S}_1\rangle$ vs $\Gamma_0$ 
curves for small enough $t_{01}$ values. 

We anticipate that our results could guide the study of related DQD systems. 
For example, when the QDs are in a parallel geometry~\cite{Zitko2006b}, 
with both QDs connected to the leads (in that case, QD$_1$ connects through $\Gamma_1$ 
to the Fermi sea, while in T-geometry $\Gamma_1=0$). In that system, 
there is the possibility of a singlet/triplet QPT~\cite{Zarand2006b,Zitko2012,Liao2015} 
through variation of the coupling parameters. For instance, when increasing 
$\Gamma_1$ (for fixed $\Gamma_0$ and $t_{01}$), the system eventually goes 
through a singlet/triplet QPT (as shown in Fig.~7(a) in 
Ref.~\cite{Liao2015})~\cite{notation}. It would be important to distinguish what 
is the character of the singlet-side of the QPT, viz., 
the TSK, the molecular, or the crossover between them. Our results 
may provide guidance to experimentalists to better place the system in 
the desired regime. 

Thus, we trust that our analysis allows experimental groups to tune the parameters of a 
DQD system with T-shape geometry into the \emph{true} TSK regime. Varying parameters 
accordingly, they could study the crossover to the molecular regime, and turn on 
$\Gamma_1$ to study the singlet/triplet QPT starting from different singlet regimes. 

\begin{acknowledgments}
We thank R.~\v{Z}itko for helping with certain details of the use of the 
NRG Ljubljana package. P.A.A., S.E.U. and G.B.M. acknowledge support from 
the CAPES-PrInt/UFU program. P.A.A. thanks the Brazilian funding agency CAPES 
for financial support. E.V. acknowledges financial support from the National 
Council for Scientific and Technological Development (CNPq), Grant No. 
311366/2021-0. S.E.U. acknowledges support from the US Department of Energy, 
Office of Basic Energy Sciences, Materials Science and Engineering Division. 
E.V.A. acknowledges financial support from the National 
Council for Scientific and Technological Development (CNPq), Grant No. 
307644/2022-7. 

\end{acknowledgments}

\bibliography{TSK}

\begin{thebibliography}{56}
\bibitem [{nor()}]{normalization}%
  \BibitemOpen
  \href@noop {} {}\bibinfo {note} {Note that, as expected, $A_{\rm eff} (\omega)$ obbeys the Friedel Sum Rule, i.e., $\pi \Gamma_{\rm eff} A_{\rm eff} (0)=1$.}\BibitemShut {Stop}%
\end{thebibliography}

\end{document}